\documentclass[preprintnumbers,amsmath,amssymb,floatfix,11pt,prd,onecolumn,
superscriptaddress,nofootinbib]{revtex4}
\usepackage{latexsym}
\usepackage{epsfig}
\usepackage{epstopdf}
\usepackage{amssymb}

\begin{document}

\title{\bf Thermodynamic Relations for Kiselev and Dilaton Black Hole
 }
\author{Bushra Majeed}
\email{bushra.majeed@sns.nust.edu.pk}
\affiliation{School of Natural
Sciences (SNS), National University of Sciences and Technology
(NUST), H-12, Islamabad, Pakistan}

\author{Mubasher Jamil}
\email{mjamil@sns.nust.edu.pk}\affiliation{School of Natural Sciences (SNS),
National University of Sciences and Technology (NUST), H-12,
Islamabad, Pakistan}
\author{Parthapratim Pradhan}
\email{pppradhan77@gmail.com}\affiliation{ Department of Physics, Vivekananda Satavarshiki Mahavidyalaya,
(Affiliated to Vidyasagar University), Manikpara, Jhargram,
West Midnapur, West Bengal 721513, India}

\begin{abstract}
{\bf Abstract:} We have studied some thermodynamics features of Kiselev black hole and dilaton black hole. Specifically we
consider Reissner Nordstr\"{o}m black hole surrounded by
radiation and dust, and Schwarzschild black hole surrounded by quintessence, as special cases of Kiselev solution.
We have calculated the products of black hole thermodynamics parameters, including surface gravities, surface temperatures, Komar energies,
areas, entropies, horizon radii and the irreducible masses, at the
inner and outer horizons. The products of
surface gravities, surface temperature product and product of
Komar energies at the horizons are not universal quantities.
For Kiselev solutions products of areas and
entropies at both the horizons are independent of mass of the black holes (except for Schwarzschild black hole surrounded by quintessence).
 For charged dilaton black hole, all the
products vanish. Using the Smarr formula approach, the first law of thermodynamics is also verified for Kiselev solutions. The phase transitions in the heat capacities are also observed.
\end{abstract}
\maketitle
\newpage
\section{Introduction}
Black holes are the most exotic objects in physics and their connection with thermodynamics is
even more surprising. Just like other thermodynamical systems, black holes have physical temperature and entropy. The analogy between the black hole thermodynamics
and the four laws of thermodynamics was first proposed in 1970's \cite{ch9670, ch5271, pe7771, be3373} and the temperature ($T$) and entropy $(S)$ are analogous of the
surface gravity $(\kappa)$ and area $(A)$ of the black hole event horizon respectively. Laws of black hole thermodynamics are studied in literature \cite{ja6111}. In \cite{ca0812} universal properties of black holes and the first law of black hole inner mechanics is discussed. In \cite{wa3114} horizon entropy sums in A(dS) spacetimes is studied. In \cite{cu6279} authors have discussed the spin entropy of a rotating black hole. The study of phase transition in black
holes is a fascinating topic \cite{pa9590, ne1865}. If a black hole has a Cauchy horizon (${\cal H}^-$) and
an event horizon(${\cal H}^+$) then it is quite interesting to study different quantities like the product
of areas of a black hole on these horizons.

Products of thermal quantities of the rotating black holes \cite{cv0111, pr8714, prarx14, bushra} and  area products for stationary black hole horizons \cite{vi1413} have been studied in literature.
Calculations show that sometimes these products do not depend on the ADM (Arnowitt-Deser-Misner) mass parameter but only on the charge and angular momentum. The  relations that are independent of the
black hole mass are of particular interest because these may turn out to be “universal” and  hold
for more general solutions with nontrivial surroundings too.

 Kiselev \cite{ki8703} considered Einstein's field equation surrounded by quintessential matter and
proposed new solutions, dependent on state parameter $\omega$ of the matter surrounding black hole. Recently some dynamical aspects, i.e. collision between particles and their escape energies after collision around Kiselev black hole \cite{ja2415} have been studied. In this
work we consider the solution of Reissner Nordstr\"{o}m(RN) black hole surrounded by energy-matter, derived
by Kiselev and study the important thermodynamic features of black hole at both the  horizons of the black hole. We also consider solution of Schwarzschild black hole
surrounded by energy-matter and analyzed its different thermodynamic products. Furthermore, we have considered
the charged dilaton black hole and computed its various thermodynamic products.

The plan of the work is as follows:
In section (II), we discuss the basic aspects of RN black hole surrounded by radiation.
Results show that the products of area and entropy calculated at $\mathcal{H}^{\pm}$ are independent of mass of the black hole, while the other products are mass dependent. In subsections of (II), the  first law of thermodynamics is obtained by the Smarr formula approach, later the rest mass is written in terms of irreducible mass of the black hole, also the phase transition in heat capacity of the black hole is discussed. Section (III) consists of discussions on  thermodynamic aspects  of RN black hole surrounded by dust. In section (IV), thermodynamics of the Schwarzschild black hole surrounded by  quintessence is studied.
In section (V) we have computed the
thermodynamic product relations for dilaton black hole. All the work is concluded in the last section. We set $G = \hbar = c = 1$, throughout the calculations.

\section{RN Black Hole Surrounded by Radiation}
The spherically symmetric and static solutions for Einstein's field equations, surrounded by energy-matter,
 as investigated by
Kiselev \cite{ki8703} can be written as:
\begin{eqnarray}\label{M1}
ds^2&= &-f(r)dt^2 + \frac{1}{f(r)}dr^2+r^2(d\theta^2+ \sin^2\theta d\phi^2),
\end{eqnarray}
where
\begin{equation} \label{M01}
f(r)= 1-\frac{2{\cal M}}{r}+ \frac{Q^2}{r^2} -\frac{\sigma}{r^{3\omega+1}},
\end{equation}
 here ${\cal M}$ and $Q$ are the mass and electric charge, of the black hole respectively,  $\sigma$ is the normalization parameter
and $\omega$ is the state
parameter of the matter around black hole. We consider the cases when RN black hole is surrounded by
radiation ($\omega= 1/3$) and dust ($\omega= 0$). For $\omega= 1/3$ two horizons
of the black hole are obtained from:
\begin{equation}
1-\frac{2{\cal M}}{r}+ \frac{Q^2}{r^2} -\frac{\sigma_r}{r^2}=0,
\end{equation}i.e.
\begin{equation} \label{M3}
r_{\pm}={\cal M}\pm \sqrt{{\cal M}^2 -Q^2 + \sigma_r}.
\end{equation}
Here $\sigma_r$ denotes the normalization parameter for radiation case, with dimensions, $[\sigma_r]=L^2$, where $L$ denotes length, $r_+$ is the outer horizon named as event horizon
${\cal H}^+$ and $r_-$ is the inner horizon known as Cauchy
horizon ${\cal H}^{-}$, ${\cal H}^{\pm}$ are the null surfaces
of infinite blue-shift and infinite red-shift respectively \cite{ch83}. Using Eq. (\ref{M3}) one can obtain,
\begin{equation}\label{M4}
r_+ r_- = Q^2 -\sigma_r,
\end{equation}
so product of horizons is independent of mass of the black hole but depends on electric charge and $\sigma_r$.
Areas of both horizons of the black hole are:
\begin{equation}\label{M5}
\mathcal{A_{\pm}}= \int^{2\pi}_0\int^\pi_0 \sqrt{g_{\theta\theta} g_{\phi \phi}}d\theta d\phi=4 \pi r_{\pm}^2= 4\pi(2{\cal M}r_{\pm} -Q^2 +\sigma_r).
\end{equation}
The corresponding semi-classical Bakenstein-Hawking entropy at ${\cal H}^{\pm}$ is \cite{ha4471}:
\begin{eqnarray}\label{M7}
\mathcal{S}_{\pm}&=&\frac{\mathcal{A}_{\pm}}{4}\nonumber\\
&= &\pi(2{\cal M}r_{\pm} -Q^2 +\sigma_r).
\end{eqnarray}
Hawking temperature of ${\cal H}^{\pm}$ is determined by using the formula
\begin{eqnarray}\label{M9}
T_{\pm}&=&\frac{1}{4 \pi}\frac{df}{dr}\mid_{r=r_{\pm}}\nonumber\\
&=&\frac{1}{4 \pi}\Big[\frac{r_{\pm}^2-Q^2+\sigma_r}{r_{\pm}^3}\Big] \nonumber\\
&=&\frac{r_{\pm}- {\cal M}}{2\pi(2{\cal M}r_{\pm} -Q^2 +\sigma_r)}.
\end{eqnarray}
    Surface gravity is the force required to an observer at infinity, for holding a particle in place, which is equal to the acceleration at horizon due to gravity of a black hole \cite{po02}:
\begin{equation}\label{M8}\kappa_{\pm}=\frac{1}{2}\frac{df}{dr}\mid_{r=r_{\pm}}=2\pi T_{\pm},\end{equation}
\begin{equation}\kappa_{\pm}=\frac{r_{\pm}- {\cal M}}{(2{\cal M}r_{\pm} -Q^2 +\sigma_r)}.
\end{equation}
The Komar energy of the black hole is defined as \cite{ko3459}
\begin{equation}
E_{\pm}= 2 \mathcal{S}_{\pm} T_{\pm}={ r_{\pm}-{\cal M}}.\label{M10}
\end{equation}
 Products of surface gravities and surface temperatures at ${\cal H}^{\pm}$ are
\begin{equation}
\kappa_+\kappa_-=4\pi^2 T_+ T_-= \frac{Q^2-\sigma_r-{\cal M}^2}{(Q^2-\sigma_r)^2}.\label{M11}
\end{equation}
The Komar energies at ${\cal H}^{\pm}$ results in
\begin{equation}\label{M13}
E_+E_-= {Q^2-{\cal M}^2- \sigma_r}.
\end{equation}
Since these products (except product of horizons $r_+r_-$) are mass dependent, so are not universal quantities. The products of areas
and entropies at ${\cal H}^{\pm}$, \begin{equation}\mathcal{A}_+\mathcal{A}_-=16 \mathcal{S}_+\mathcal{S}_-=16 \pi^2(Q^2-\sigma_r)^2,\label{area}
\end{equation}
 are mass independent, so these are  the universal parameters of the black hole.
\subsection{Smarr Formula for Cauchy Horizon (${\cal H}^-)$}
The expression of the area of black hole is \cite{sm7173}:
\begin{equation}\label{M14}
\mathcal{A}=4 \pi \Big[ 2M^2 -Q^2 +\sigma_r+2M\sqrt{M^2-Q^2+\sigma_r}\Big],
\end{equation} from the area of both the horizons
\begin{equation}\label{M15}
\mathcal{A}_{\pm}=4 \pi \Big[ 2M^2 -Q^2 +\sigma_r\pm 2M\sqrt{M^2-Q^2+\sigma_r}],
\end{equation}
 one can write the ADM mass of the black hole as:
\begin{eqnarray}\label{M16}
{\cal M}^2&=& \frac{\mathcal{A}_{\pm}}{16 \pi}+\frac{Q^4 \pi }{\mathcal{A}_{\pm}}
+\frac{Q^2}{2}- \frac{2\pi \sigma_r Q^2}{\mathcal{A}_{\pm}}-\frac{\sigma_r}{2}+
\frac{\pi \sigma_r^2}{\mathcal{A}_{\pm}}\nonumber\\
&=&\frac{\mathcal{A}_{\pm}}{16 \pi}+\frac{\pi Q^4}{\mathcal{A}_{\pm}}
+{Q^2}\Big(\frac{\mathcal{A}_{\pm}-4 \pi \sigma_r}{2\mathcal{A}_{\pm}}\Big)-
\sigma_r \Big(\frac{\mathcal{A}_{\pm}-2\pi \sigma_r}{2 \mathcal{A}_{\pm}}\Big).
\end{eqnarray}
According to the first law of thermodynamics, differential of mass of the black hole can be related with the  change in its area and electric charge. Since  the effective surface tension at the horizon is proportional to the temperature of the black hole horizons, so we can write:
\begin{equation}\label{M17}
d{\cal M}=\mathcal{T_{\pm}} d\mathcal{A}_{\pm} +
\Phi_{\pm}dQ,
\end{equation} where $\mathcal{T}_{\pm}$ and $\Phi_{\pm}$ are defined as:
\begin{eqnarray}\label{M18} \mathcal{T}_{\pm}&=& \text{Effective surface tension at horizons}\nonumber\\
&=&\frac{1}{{\cal M}}\Big(\frac{1}{32\pi}-\frac{Q^4 \pi}{2 \mathcal{A}^2_{\pm}}
+\frac{Q^2 \sigma_r \pi}{\mathcal{A}^2_{\pm}}-\frac{\pi \sigma_r^2}{2 \mathcal{A}_{\pm}}\Big)\nonumber\\
\Phi_{\pm}&=& \text{Electromagnetic potentials at horizons}\nonumber\\
&=&\frac{1}{{\cal M}}\Big(\frac{2\pi Q^3}{\mathcal{A}_{\pm}}-
\frac{2\pi Q \sigma_r}{\mathcal{A}_{\pm}}+ \frac{Q}{2}\Big).
\end{eqnarray}
The effective surface tension can be rewritten as:
\begin{eqnarray}
\mathcal{T}_{\pm}&=& \frac{1}{32 \pi{\cal M}}\Big(1-\frac{16 \pi^2 (Q^4 -2Q^2\sigma_r
+\sigma_r^2)}{ \mathcal{A}^2_{\pm}}\Big)\nonumber\\&=&\frac{1}{16 \pi{\cal M}}
\Big(1-\frac{2M^2 -Q^2+\sigma_r}{ r^2_{\pm}}\Big)\nonumber\\
&=&\frac{1}{8\pi}\Big(\frac{r_{\pm}-{\cal M}}{2{\cal M}r_{\pm}-Q^2+\sigma_r}\Big)\nonumber\\
&=&\frac{\kappa_{\pm}}{8\pi}.
\end{eqnarray}
 So RN black hole surrounded by radiation satisfies the first law of thermodynamics, verified by Smarr formula approach.
\subsection{Christodoulou-Ruffini Mass Formula for RN Black Hole surrounded by Radiation}
Christodoulou and  Christodoulou and Ruffini  \cite{ch9670,ch5271} had shown that the mass of a black hole could be increased or decreased
but the irreducible mass
${\cal M}_{\text{irr}}$ of a black hole can not be decreased. In fact , most processes
result in an increase in ${\cal M}_{\text{irr}}$ and during reversible process this quantity also does
not change. Also, the surface area of a black hole has behavior \cite{ba6173}
 \begin{eqnarray}
 d{\cal A}_{\pm} \geq 0 \label{arth},
\end{eqnarray} so there exist a relation between area and irreducible
mass. The ${\cal M}_{\text{irr}}$ is proportional to the square root of
the black hole's area. Since the RN-Radiation space-time has regular event horizon and Cauchy
horizons so the irreducible mass of a black hole is proportional to
the square root of its surface area \cite{ch5271}\begin{equation}\sqrt{\frac{\mathcal{A}_{\pm}}{16 \pi}}={\cal M}_{\text{irr}\pm}= \sqrt{\frac{r^2_{\pm}}{4}},
\end{equation}where ${\cal M}_{\text{irr}-}$ and ${\cal M}_{\text{irr}+}$
are irreducible masses defined on ${\cal H}^{\pm}$ respectively.
Product of the irreducible masses at
$\mathcal{H^{\pm}}$ is:
\begin{eqnarray}\label{M22}
{\cal M}_{\text{irr}+} {\cal M}_{\text{irr}-}&=& \sqrt{\frac{\mathcal{A}_+ \mathcal{A}_-}{(16 \pi)^2}}\nonumber\\
&=& \frac{Q^2-\sigma_r}{4}.
\end{eqnarray}
This product is universal because it does not depend on mass of the black hole.
The expression for the rest mass of the rotating
charged black hole given by Christodoulou and Ruffini in
terms of its irreducible mass, angular momentum and charge
is \cite{ch5271}:
\begin{equation}
{\cal M}^2= ({\cal M}_{\text{irr}\pm}+ \frac{Q^2}{4{\cal M}_{\text{irr}\pm}})^2
+ \frac{J^2}{4 {\cal M}^2_{\text{irr}\pm}}.\label{M19}
\end{equation}
  Setting $\rho_{\pm}= 2 {\cal M}_{\text{irr}_{\pm}}$, mass of the RN black hole surrounded by radiation becomes:
\begin{equation}{\cal M}^2=\frac{\rho^4_{\pm}+ Q^4}{4\rho ^2_{\pm}}
+\frac{\pi}{2}(\sigma^2_r-2\sigma_r Q^2)+ 2{\cal M}_{\text{irr+}} {\cal M}_{\text{irr}-}.
\end{equation}


\subsection{Heat Capacity $C_{\pm}$ on ${\cal H}^{\pm}$}
\begin{figure}
\centering
\includegraphics [width=9cm]{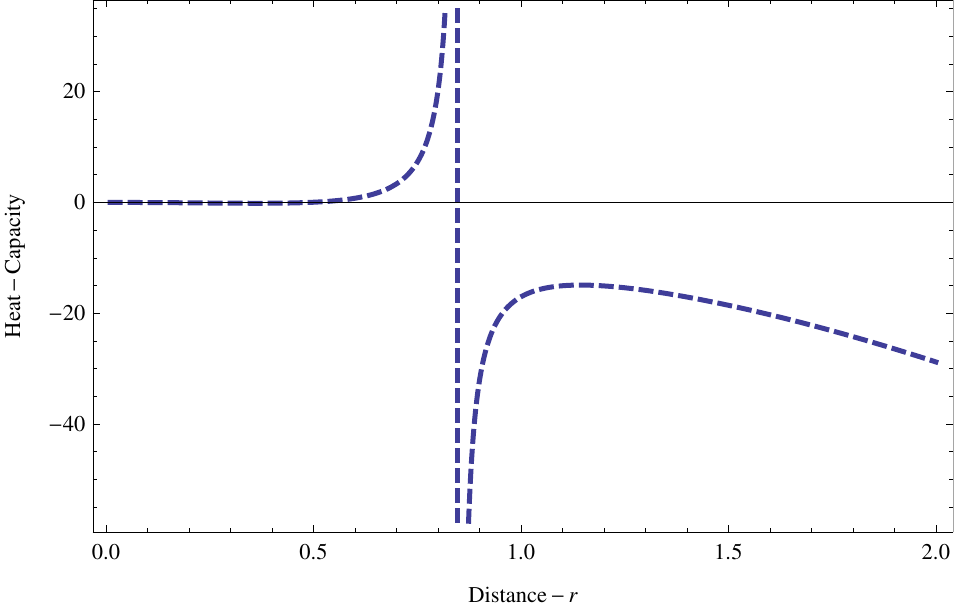}
\caption{Heat capacity undergoes phase transition from instability to stability, diverges at $r=\sqrt{3(Q^2- \sigma_r)}$ and again goes to instable region, we chose $\sigma_r=0.01$ and $Q=0.5$} \label{1M}\end{figure}

Another important measure to
study the thermodynamic properties of a black hole is the heat capacity, its nature (positivity or negativity) reflects the stability or instability  of a thermal system (black hole). A black hole with negative heat capacity is in unstable equilibrium state, i.e.  it may decay to a hot flat space by emitting Hawking radiations or it may grow without limit by absorbing  radiations \cite{gr3082}.
One can get heat capacity of a black hole by using
\begin{equation}
C_{\pm}= \frac{\partial {\cal M}}{\partial T_{\pm}},
\end{equation}
where mass ${\cal M}$ in terms of $r_{\pm}$ is:
\begin{equation}
{\cal M}=\frac{r_{\pm}^2 +Q^2 -\sigma_r}{2r_{\pm}}.
\end{equation}
The partial derivatives of mass ${\cal M}$ and temperature $T_{\pm}$ with respect to $r_{\pm}$ are:
\begin{equation}
\frac{\partial {\cal M}}{\partial r_{\pm}}=\frac{r_{\pm}^2-Q^2+\sigma_r}{2 r_{\pm}^2},
\end{equation}
and from Eq. (\ref{M9}) we have
\begin{equation}
\frac{\partial T}{\partial r_{\pm}}=\frac{1}{2 \pi}\frac{(3 Q^2-r_{\pm}^2-3\sigma_r)}{r_{\pm}^4}.
\end{equation}
The expression for heat capacity $C_{\pm}=\frac{\partial {\cal M}}{\partial r_{\pm}}
\frac{\partial r_{\pm}}{\partial T_{\pm}}$ for RN black hole surrounded by radiation at horizons becomes:
\begin{equation} \label{M21}
C_{\pm}=\frac{2 \pi r_{\pm}^2 (r_{\pm}^2 - Q^2 +\sigma_r)}{3Q^2- 3 \sigma_r-r_{\pm}^2}.
\end{equation}
Note that there are two possible cases for heat capacity to be positive:\\
\textbf{Case $1$:} When both  $r_{\pm}^2 - Q^2 +\sigma_r$ and $3Q^2- 3 \sigma_r-r_{\pm}^2$ are positive.\\
\textbf{Case $2$:} When both  $r_{\pm}^2 - Q^2 +\sigma_r$ and $3Q^2- 3 \sigma_r-r_{\pm}^2$ are negative.\\
Since we are interested in positive $r$ only, so case-$1$ implies that heat capacity is positive if
\begin{equation}\sqrt{Q^2- \sigma_r}~<r <~\sqrt{3(Q^2- \sigma_r)},\end{equation}
from case-$2$ we get \begin{equation}3(Q^2- \sigma_r)~<r^2<~(Q^2- \sigma),\end{equation}which is not possible,
so we exclude this case.\\
Heat Capacity is negative if\\
\textbf{Case $a$}:  $r_{\pm}^2 - Q^2 +\sigma_r>0$ and $3Q^2- 3 \sigma_r-r_{\pm}^2<0$,\\
\textbf{Case $b$}: $r_{\pm}^2 - Q^2 +\sigma_r<0$ and $3Q^2- 3 \sigma_r-r_{\pm}^2>0$.\\
From case-($a$) implies that for:
\begin{equation}r> \sqrt{3Q^2-3\sigma_r},\end{equation} heat capacity is negative. From case-($b$) we get negative  capacity for
\begin{equation}-\sqrt{Q^2-\sigma}~<r<~\sqrt{Q^2-\sigma},\end{equation}  we are interested in positive $r$ only.
The region where heat capacity is negative, corresponds
to an instable region around black hole, whereas a region in
which the heat capacity is positive, represents a stability
region.
Behavior of heat capacity given in Eq. (\ref{M21}) is shown in Fig. (\ref{1M}). Heat capacity is negative in the region $0<r<0.4898 $ and $r> 0.8485$, while positive for $0.4898<r<0.8485$.
Interestingly, the product of heat capacity on ${\cal H}^{\pm}$ becomes
\begin{eqnarray}\label{psh}
C_{+} C_{-} &=& 4\pi^2 \left(Q^2-\sigma_r \right)^2 \frac{\left[ (Q^2-\sigma_r)-{\cal M}^2 \right]}
{\left[4(Q^2-\sigma_r)-3{\cal M}^2\right]},
\end{eqnarray}
the product depends on mass parameter and charge parameter.
Thus the product of specific heats is not universal.
\section{RN Black Hole Surrounded by Dust}
Metric of RN black hole surrounded by dust is same as in Eq. (\ref{M1}), $f(r)$ defined in Eq. (\ref{M01})
 with $\omega= 0$ and $\sigma= \sigma_d$ becomes: \begin{equation}f(r)= 1-\frac{2{\cal M}}{r}+\frac{Q^2}{r^2}-\frac{\sigma_d}{r}\label{M001},\end{equation}
where $[\sigma_d]= L$. The horizons are
\begin{equation}\label{M03}
r_{\pm}=\frac{2{\cal M}+\sigma_d \pm \sqrt{(2{\cal M}+\sigma_d)^2-4Q^2}}{2}.
\end{equation}
Area of the horizons ${\cal H}^{\pm}$ is:
\begin{eqnarray}\label{M05}
\mathcal{A_{\pm}}&=&\pi[2(2{\cal M}+\sigma_d)^2-4Q^2\pm 2(2{\cal M}+\sigma_d)\sqrt{(2{\cal M}+\sigma_d)^2-4 Q^2}],\nonumber\\
&=& 4\pi [(2{\cal M}+\sigma_d)r_{\pm}-Q^2].
\end{eqnarray}
Entropy of the horizons is
\begin{eqnarray}\label{M07}
\mathcal{S}_{\pm}= \pi[(2{\cal M}+\sigma_d)r_{\pm}-Q^2].
\end{eqnarray}
Surface gravity and Hawking temperature of horizons are respectively:
\begin{equation}\label{M08}
\kappa_{\pm}=\frac{2r_{\pm}-(2 {\cal M}+\sigma_d)}{2[(2{\cal M}+\sigma_d)r_{\pm}-Q^2]},
\end{equation}
and
\begin{eqnarray}\label{M09}
T_{\pm}
&=& \frac{2r_{\pm}-(2 {\cal M}+\sigma_d)}{4 \pi[(2{\cal M}+\sigma_d)r_{\pm}-Q^2]},\nonumber\\
&=&\frac{1}{4 \pi}\Big(\frac{r_{\pm}^2-Q^2}{r_+^3}\Big),
\end{eqnarray} where we have used $r_{\pm}^2= (2{\cal M}+\sigma_d)r_{\pm}-Q^2$.
The Komar energy becomes:
\begin{equation}
E_{\pm}= \frac{2r_{\pm}-(2 {\cal M}+\sigma_d)}{2}.\label{M010}
\end{equation}
Product of surface gravities and temperatures at the horizons, are
\begin{equation}\kappa_+\kappa_-=4\pi^2 T_+ T_-=- \frac{(2{\cal M}+\sigma_d)^2-4Q^2}{4Q^4}.\label{dM11}
\end{equation}
Product of Komar energies at the
horizons is:
\begin{equation}\label{dM13}
E_+E_-= \frac{4Q^2-(2{\cal M}+\sigma_d)^2}{4}.
\end{equation} Note that all products are mass dependent, so these quantities are not universal.
Products of areas and entropies at both horizons ${\cal H}^{\pm}$ are:
\begin{equation}
\mathcal{A}_+\mathcal{A}_-= 16\mathcal{S}_{+}\mathcal{S}_{-}=16 \pi^2 Q^4.
\end{equation}
It is clear that area product and entropy product are universal entities.
\subsection{Smarr Formula for Cauchy Horizon (${\cal H}^-)$}
Area of both the horizons must be constant given by
\begin{eqnarray}\label{M015}
\mathcal{A}_{\pm}&=&\pi[2(2{\cal M}+\sigma_d)^2-4Q^2\pm 2(2{\cal M}+\sigma_d)\sqrt{(2{\cal M}+\sigma_d)^2-4 Q^2}].
\end{eqnarray}
Using Eq. (\ref{M015}) mass of the black hole or ADM mass is
expressed in terms of the areas of horizons as:
\begin{equation}\label{M016}
{\cal M}^2+{\cal M}\sigma_d = \frac{\mathcal{A}_{\pm}}{16 \pi}+\frac{\pi Q^4 }{\mathcal{A}_{\pm}}-\frac{(\sigma^2_d-2Q^2)}{4}.
\end{equation}
Differential of mass could be expressed in terms of physical invariants
of the horizons,
\begin{equation}\label{M017}
d{\cal M}=\mathcal{T_{\pm}} d\mathcal{A}_{\pm} +
\Phi_{\pm}dQ,
\end{equation}
where
\begin{eqnarray}\label{M018}
\mathcal{T}_{\pm}&=& \frac{1}{(2{\cal M}+\sigma_d)}\Big(\frac{1}{16 \pi}-\frac{\pi Q^4}{\mathcal{A}^2_{\pm}}\Big),\nonumber\\
\Phi_{\pm}&=&\frac{1}{(2{\cal M}+\sigma_d)}\Big(Q+\frac{4\pi Q^3}{\mathcal{A}_{\pm}}\Big).
\end{eqnarray}
We can rewrite effective surface tension as:
\begin{eqnarray}
\mathcal{T}_{\pm} &=& \frac{1}{16\pi(2{\cal M}+\sigma_d)}\Big(1-\frac{16\pi^2 Q^4}{\mathcal{A}^2_{\pm}}\Big),\nonumber\\
&=&\frac{1}{8 \pi(2{\cal M}+\sigma_d)} \Big[1-\frac{4({\cal M}^2+{\cal M}\sigma_d)+(\sigma^2_d-2Q^2)}{2 r^2_{\pm}}\Big].
\end{eqnarray} Or
\begin{eqnarray}\mathcal{T}&=&\pm \frac{\sqrt{(2\mathcal{M}+\sigma_d)^2-4Q^2}}{16 \pi ((2\mathcal{M}+\sigma_d)r_{\pm}-Q^2)},\nonumber\\
&=&\frac{\kappa_{\pm}}{8\pi}.
\end{eqnarray}
So the first law of black hole thermodynamics is
verified, for RN black hole surrounded by dust, using the Smarr formula approach.
\subsection{Christodoulou-Ruffini Mass Formula for RN Black Hole Surrounded by Dust}
The expression for irreducible mass for
RN black hole surrounded by dust is
\begin{equation}{\cal M}_{\text{irr}\pm}= \frac{2{\cal M}+\sigma_d \pm \sqrt{(2{\cal M}+\sigma_d)^2-4Q^2}}{4}.
\end{equation}
Here
${\cal M}_{\text{irr}-}$ and ${\cal M}_{\text{irr}+}$  are irreducible masses defined on inner and outer horizons respectively.
Area of ${\cal H}^{\pm}$, in terms of ${\cal M}_{\text{irr }\pm}$ is:
\begin{equation}\label{M20}
\mathcal{A_{\pm}}= 16 \pi ({\cal M}_{\text{irr}\pm})^2.
\end{equation}
Product of the irreducible mass at the horizons
$\mathcal{H^{\pm}}$ is:
\begin{eqnarray}\label{M22}
{\cal M}_{\text{irr}+} {\cal M}_{\text{irr}-}&=& \sqrt{\frac{\mathcal{A}_+ \mathcal{A}_-}{(16 \pi)^2}}\nonumber\\
&=& \frac{Q^2}{4}.
\end{eqnarray}
This product is independent of mass of the black hole.
Mass of the black hole expressed in terms of its irreducible mass and charge is:
\begin{equation}{\cal M}^2+ \mathcal{M}\sigma_d=\frac{\rho^4_{\pm}+ Q^4}{4\rho ^2_{\pm}}
-\sigma_d +2{\cal M}_{\text{irr+}} {\cal M}_{\text{irr}-}.
\end{equation}
\subsection{Heat Capacity $C_{\pm}$ on ${\cal H}^{\pm}$}
\begin{figure}
\centering
\includegraphics [width=9cm]{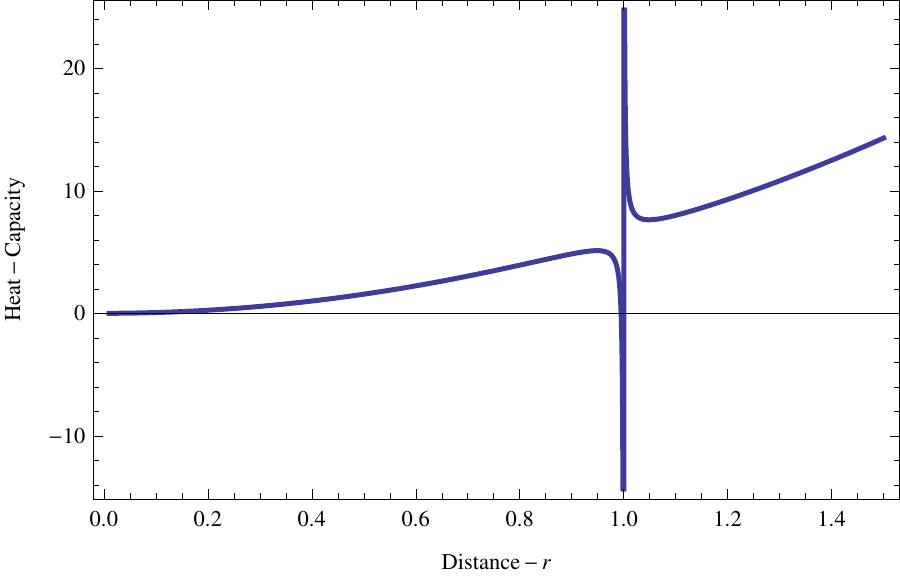}\caption{Heat capacity undergoes phase transition from stability to
instability region, we chose $\sigma_d=0.01$, $Q=0.5$ and ${\cal M}=1$.}\label{2M}\end{figure}

Mass of RN black hole surrounded by dust in terms of $r_{\pm}$ is:
\begin{equation}
{\cal M}=\frac{r_{\pm}^2 +Q^2 -\sigma_d r_{\pm}}{2r_{\pm}}.
\end{equation}
Partial derivatives of mass ${\cal M}$ and temperature $T_{\pm}$ with respect to $r_{\pm}$ are:
\begin{equation}
 \frac{\partial {\cal M}}{\partial r_{\pm}}=\frac{r_{\pm}^2-Q^2+\sigma_d r_{\pm}}{2 r_{\pm}^2},
 \end{equation} and
 \begin{equation}
 \frac{\partial T_{\pm}}{\partial r_{\pm}}=\frac{1}{4 \pi}\frac{(3 Q^2-r_{\pm}^2)}{r_{\pm}^4},
 \end{equation}where $T$ is given in Eq. (\ref{M09}).
 The heat capacity $C=\frac{\partial {\cal M}}{\partial r_{\pm}}\frac{\partial r_{\pm}}{\partial T}$
 at the horizons is:
\begin{equation} \label{M23}
C_{\pm}=\frac{2 \pi r_{\pm}^2 (r_{\pm}^2 - Q^2 +\sigma_d r_{\pm})}{r_{\pm}^2-Q^2}.
\end{equation}
In this case the product formula for heat capacity is found to be
\begin{eqnarray}\label{psh1}
C_{+} C_{-} &=& \frac{4\pi^2 Q^4 \left[ 4Q^4-Q^2(2{\cal M}+\sigma_d)^2+\sigma_d^2 Q^2 \right]}{4Q^4-Q^2 (2{\cal M}+\sigma_d)^2}.
\end{eqnarray}
It is clear that the product formula does depend on mass parameter, so it is not universal in nature.
Note that there are two possible cases for heat capacity, $C$, to be positive:\\
\textbf{Case $1$:} When both  $r_{\pm}^2 - Q^2 +\sigma_d r_{\pm}$ and $r^2-Q^2$ are positive.\\
\textbf{Case $2$:} When both $r_{\pm}^2 - Q^2 +\sigma_d r_{\pm}$ and $r^2-Q^2$ are negative.\\
  Considering $\sigma_dr_{\pm}$ as a positive quantity (for physically accepted region, $r$) from case-1, we can say $C$ is positive for only $r_{\pm}^2 - Q^2 >0$ i.e.
\begin{equation}\sqrt{Q^2}<~r~<-\sqrt{Q^2},\end{equation}
while from case-$2$ we can say $C$ is negative for only $r_{\pm}^2 - Q^2 +\sigma_d r_{\pm}<0$ i.e.  \begin{equation}r~< \frac{-\sigma_d}{2}+\frac{1}{2}\sqrt{4Q^2+ \sigma_d^2}.\end{equation}
Heat capacity is negative if\\
\textbf{Case $a$}:  $r_{\pm}^2 - Q^2 +\sigma_d r_{\pm}>0$ and $r_{\pm}^2-Q^2<0$,\\
\textbf{Case $b$}: $r_{\pm}^2 - Q^2 +\sigma_d r_{\pm}<0$ and $r_{\pm}^2-Q^2>0$.\\
Case ($b$) is not possible mathematically since $\sigma_d r_{\pm}>0$, while in case ($a$) heat capacity is negative in the region, where $r$ satisfies both the conditions
\begin{equation}\frac{-\sigma_d}{2}+\frac{1}{2}\sqrt{4Q^2+ \sigma_d^2}~<r<~\frac{-\sigma_d}{2}-\frac{1}{2}\sqrt{4Q^2+ \sigma_d^2},\end{equation} and
\begin{equation}-\sqrt{Q^2}~<~r<~\sqrt{Q^2},\end{equation} simultaneously.
We consider $\sigma$ and $Q$ both are positive in all the calculations.
Behavior of the heat capacity given in Eq. (\ref{M23}) is shown in Fig. (\ref{2M}).
\section{Schwarzschild Black Hole Surrounded by Quintessence}
Metric for Schwarzschild black hole surrounded by quintessence is same as defined in Eq. (\ref{M1}) and $f(r)$ defined in Eq. (\ref{M01}) with $\omega= -2/3$, $\sigma=\sigma_q$ and $Q=0$ becomes:
\begin{equation}\label{M0001}f(r)=1-\frac{2{\cal M}}{r} -\sigma_q r, \end{equation} where dimensions of $\sigma_q$ are that of $L^{-1}$. The horizons, $r_{\pm}$,
of the black hole are:
\begin{equation} \label{oM3}
r_{\pm}=\frac{1\pm \sqrt{1-8{\cal M}\sigma_q}}{2\sigma_q}.
\end{equation}
 Product of the two horizons yield:
\begin{equation}\label{sM4}
r_+ r_- = \frac{2{\cal M}}{\sigma_q},
\end{equation} and it is depending on mass of the black hole and $\sigma_q$.
Areas of the horizons are:
\begin{eqnarray}\label{oM5}\mathcal{A_{\pm}}=4\pi\Big[\frac{r_{\pm}-2{\cal M}}{\sigma_q}\Big].
\end{eqnarray}
Entropy at the horizons ${\cal H}^{\pm}$ is:
\begin{eqnarray}\label{oM7}\mathcal{S}_{\pm}&= &\frac{\pi}{\sigma_q}(r_{\pm} -2{\cal M}).
\end{eqnarray}
Hawking temperature of the horizons is:
\begin{eqnarray}\label{oM9}
T_{\pm}&=& \frac{1}{4 \pi}\Big[\frac{1-2\sigma_q r_{\pm}}{r_{\pm}}\Big],
\end{eqnarray} and
the surface gravity on the black hole horizons ${\cal H}^{\pm}$ is given by:
\begin{equation}\label{oM8}
\kappa_{\pm}=\frac{1}{2}\Big[\frac{1-2\sigma_q r_{\pm}}{r_{\pm}}\Big].
\end{equation}
The Komar energy is given by:
\begin{equation}E_{\pm}= \frac{2\mathcal{M}(1+2\sigma_qr_{\pm})-r_{\pm}}{2\sigma_q r_{\pm}}.\label{oM10}
\end{equation}
Product of surface gravities and temperatures of ${\cal H}^{\pm}$ is:
\begin{equation}\kappa_+ \kappa_-=4\pi^2T_+ T_-=\frac{(8 \mathcal{M} \sigma_q-1)\sigma_q}{8 \mathcal{M}}.\label{oM11}
\end{equation}
Product of Komar energies of the horizons is:
\begin{equation}\label{oM13}
E_+E_-= \frac{\mathcal{M}(8 \mathcal{M} \sigma_q-1)}{2\sigma_q},
\end{equation}
respectively. It is clear that all these products are depending on
mass of the black hole, so these quantities are not universal. The products of areas and entropies at ${\cal H}^{\pm}$ are:
\begin{equation}
\mathcal{A}_+\mathcal{A}_-=16\mathcal{S}_+\mathcal{S}_-= \Big(\frac{8 \pi{\cal M}}{\sigma_q}\Big)^2,
\end{equation}
again both products are not universal quantities.
\subsection{Smarr Formula for Cauchy Horizon (${\cal H}^-)$}
Writing area of both horizons of the black hole as:
\begin{equation}\label{oM15}
\mathcal{A}_{\pm}=\frac{\pi}{\sigma^2_q} \Big[ 2-8{\cal M}\sigma_q \pm 2\sqrt{1-8{\cal M}\sigma_q}\Big].
\end{equation}
Using Eq. (\ref{oM15}) mass of the black hole or ADM mass is
expressed in terms of the areas of horizons as:
\begin{eqnarray}\label{oM16} 4{\cal M}^2+\frac{{\cal M}\mathcal{A}\sigma_q}{\pi} &=&
\frac{1}{16 \pi}\Big[4\mathcal{A}_{\pm}-\frac{\mathcal{A}_{\pm}^2\sigma^2_q}{\pi}\Big].
 \end{eqnarray}
Differential of mass, expressed in terms of physical invariants
of the horizons is: \begin{equation}\label{sM17}
d{\cal M}=\mathcal{T_{\pm}} d\mathcal{A}_{\pm},\end{equation} where
\begin{eqnarray}\label{sM18}
\mathcal{T}_{\pm} &=& \frac{1}{8\pi {\cal M}+\sigma_q \mathcal{A}_{\pm}}
                  \Big[\frac{4\pi-16\pi\sigma_q{\cal M}-2\mathcal{A}_{\pm} \sigma^2_q}{16 \pi}\Big]\nonumber\\
&=&\frac{1}{16 \pi \mathcal{A}_{\pm}}[8 \pi \mathcal{M}- \mathcal{A}_{\pm}\sigma_q]
                  \nonumber\\
                 &=& \frac{\kappa_{\pm}}{8\pi},
\end{eqnarray}
where we have used $\mathcal{M}= (r_{\pm}-\sigma_q r_{\pm}^2)/2$, and $\kappa $ is defined in Eq. (\ref{oM8}). Hence the first law of thermodynamics is satisfied by Schwarzschild black hole surrounded by quintessence.
\subsection{Christodoulou-Ruffini Mass Formula for Schwarzschild Black Hole Surrounded by Quintessence}
The irreducible mass of
Schwarzschild black hole surrounded by quintessence is
\begin{equation}{\cal M}_{\text{irr}\pm}= \frac{1\pm \sqrt{1-8{\cal M}\sigma_q}}{4\sigma_q}.
\end{equation}
Here
${\cal M}_{\text{irr}-}$ and ${\cal M}_{\text{irr}+}$  are irreducible masses defined on inner and outer horizons respectively.
Area of ${\cal H}^{\pm}$, in terms of ${\cal M}_{\text{irr }\pm}$ is:
\begin{equation}\label{M20}
\mathcal{A_{\pm}}= 16 \pi ({\cal M}_{\text{irr}\pm})^2.
\end{equation}
Product of the irreducible mass at the horizons
$\mathcal{H^{\pm}}$ is:
\begin{eqnarray}\label{M22}
{\cal M}_{\text{irr}+} {\cal M}_{\text{irr}-}&=& \sqrt{\frac{\mathcal{A}_+ \mathcal{A}_-}{(16 \pi)^2}}\nonumber\\
&=& \frac{\mathcal{M}}{2\sigma}.
\end{eqnarray}
This product is depending on mass of the black hole.
Expression of mass given in Eq. (\ref{oM16}), in terms of irreducible mass becomes:
\begin{equation} 8 \sigma^2 {\cal M}^2_{\text{irr}_{+}} {\cal M}^2_{\text{irr}_{-}}+4\pi \rho^2_{\pm}({\cal M}_{\text{irr}_{+}}{\cal M}_{\text{irr}_{-}})=4 \rho^2_{\pm}(1-4\sigma^2).
\end{equation}
\subsection{Heat Capacity $C_{\pm}$ on ${\cal H}^{\pm}$}

Mass of Schwarzschild black hole surrounded by quintessence in terms of $r_{\pm}$ is:
\begin{equation} {\cal M}=\frac{r_{\pm} -\sigma_q r_{\pm}^2}{2}.
\end{equation}
 The partial derivative of mass ${\cal M}$ with respect to $r_{\pm}$ is:
 \begin{equation}\frac{\partial {\cal M}}{\partial r_{\pm}}=\frac{1-2 \sigma_q r_{\pm}}{2},
 \end{equation}
 and using Eq. (\ref{oM9}) we get
 \begin{equation}
 \frac{\partial T_{\pm}}{\partial r_{\pm}}=-\frac{1}{4\pi}\Big[\frac{\sigma_q}{ r_{\pm}^2-2 \mathcal{M}}\Big].
 \end{equation}
 The expression for heat capacity $C=\frac{\partial {\cal M}}{\partial r_{\pm}}\frac{\partial r_{\pm}}{\partial T}$ at
 the horizon becomes:
\begin{equation} \label{sM23}
C_{\pm}=\frac{- 2\pi (1-2\sigma_q r_{\pm})(r_{\pm}-2\mathcal{M})}{\sigma_q}.
\end{equation}

\begin{figure}
\centering
\includegraphics [width=9cm]{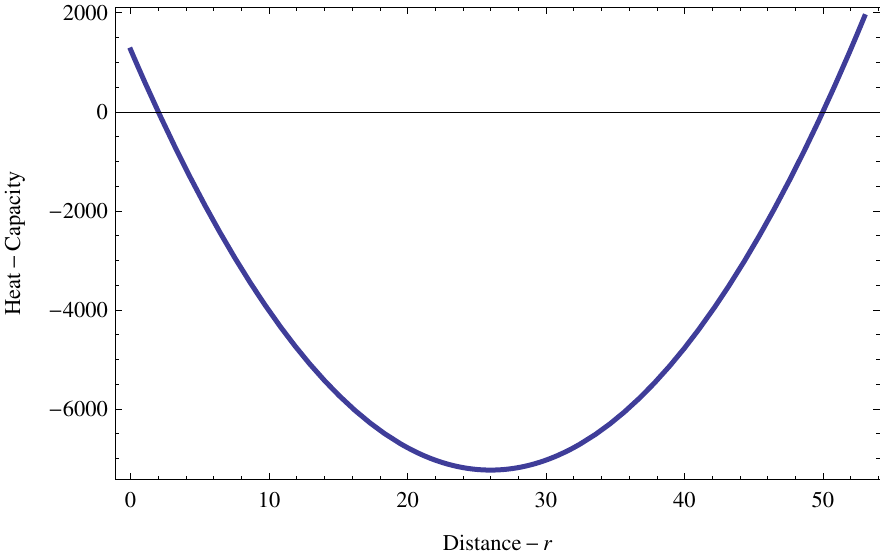}\caption{Heat capacity phase transition of Schwarzschild black hole surrounded by quintessence. We chose $\sigma_q=0.01$, and ${\cal M}=1$. Heat capacity is positive for $0<r<2$ and $r>50$, negative for $2<r<50$ and is zero at $r=2$ and $r= 50$.}\label{3M}\end{figure}

\begin{table}
  \centering
  \begin{tabular}{|c|c|c|c|}
     \hline
     Parameter & RN-Radiation & RN-Dust & Schwarzschild-Quintessence \\
     \hline
     $r_{\pm}$ & $\mathcal{M}\pm \sqrt{\mathcal{M}^-Q^2+\sigma_r}$ & $\frac{2{\cal M}+\sigma_d \pm \sqrt{(2{\cal M}+\sigma_d)^2-4Q^2}}{2}$ & $\frac{1\pm \sqrt{1-8{\cal M}\sigma_q}}{2\sigma_q}$\\

     $\mathcal{A}_{\pm}$ & $4\pi(2\mathcal{M}r_{\pm}-Q^2+\sigma_r)$ & $4\pi [(2{\cal M}+\sigma_d)r_{\pm}-Q^2] $ & $4\pi\Big[\frac{r_{\pm}-2{\cal M}}{\sigma_q}\Big]$ \\

     $S_{\pm}$ & $\pi(2\mathcal{M}r_{\pm}-Q^2+\sigma_r$ & $\pi [(2{\cal M}+\sigma_d)r_{\pm}-Q^2]$ & $\frac{\pi}{\sigma_q}(r_{\pm} -2{\cal M})$\\

     $T_{\pm}$ & $\frac{r_{\pm}- {\cal M}}{2\pi(2{\cal M}r_{\pm} -Q^2 +\sigma_r)}$ & $\frac{2r_{\pm}-(2 {\cal M}+\sigma_d)}{4 \pi[(2{\cal M}+\sigma_d)r_{\pm}-Q^2]}$ & $\frac{1}{4 \pi}\Big[\frac{1-2\sigma_q r_{\pm}}{r_{\pm}}\Big]$\\

     $\kappa_{\pm}$ & $\frac{r_{\pm}- {\cal M}}{(2{\cal M}r_{\pm} -Q^2 +\sigma_r)}$ & $\frac{2r_{\pm}-(2 {\cal M}+\sigma_d)}{2[(2{\cal M}+\sigma_d)r_{\pm}-Q^2]}$ & $\frac{1}{2}\Big[\frac{1-2\sigma_q r_{\pm}}{r_{\pm}}\Big]$ \\

     $E_{\pm}$ &$ { r_{\pm}-{\cal M}}$ & $\frac{2r_{\pm}-(2 {\cal M}+\sigma_d)}{2}$ & $\frac{2\mathcal{M}(1+2\sigma_qr_{\pm})-r_{\pm}}{2\sigma_q r_{\pm}}$\\

     $\kappa_+\kappa_-$ &$ \frac{Q^2-\sigma_r-{\cal M}^2}{(Q^2-\sigma_r)^2}$ & $- \frac{(2{\cal M}+\sigma_d)^2-4Q^2}{4Q^4}$ & $\frac{(8 \mathcal{M} \sigma_q-1)\sigma_q}{8 \mathcal{M}}$ \\
    $T_+ T_-$& $\frac{Q^2-{\cal M}^2-\sigma_r}{4\pi^2 (Q^2-\sigma_r)^2} $& $\frac{4Q^2-(2{\cal M}+\sigma_d)^2}{16 \pi ^2Q^4}$ & $\frac{(8 \mathcal{M} \sigma_q-1)\sigma_q}{32\pi^2 \mathcal{M}}$ \\

     $E_+E_-$ & ${Q^2-{\cal M}^2- \sigma_r}$ & $\frac{4Q^2-(2{\cal M}+\sigma_d)^2}{4}$ & $\frac{\mathcal{M}(8 \mathcal{M} \sigma_q-1)}{2\sigma_q}$\\

      $\mathcal{A}_+\mathcal{A}_- $&$ 16 \pi^2(Q^2-\sigma_r)^2$ & $16 \pi^2 Q^4$ & $\Big(\frac{8 \pi{\cal M}}{\sigma_q}\Big)^2$ \\

       $\mathcal{S}_+\mathcal{S}_-$& $\pi^2(Q^2-\sigma_r)^2 $& $\pi^2Q^4$ & $\Big(\frac{2\pi {\cal M}}{\sigma_q}\Big)^2$\\

       ${\cal M}_{\text{irr}+} {\cal M}_{\text{irr}-}$ & $\frac{Q^2-\sigma_r}{4}$ & $\frac{Q^2}{4}$ & $\frac{\mathcal{M}}{2\sigma}$ \\

       $C_{\pm}$ & $\frac{2 \pi r_{\pm}^2 (r_{\pm}^2 - Q^2 +\sigma_r)}{3Q^2- 3 \sigma_r-r_{\pm}^2}$ &$\frac{2 \pi r_{\pm}^2 (r_{\pm}^2 - Q^2 +\sigma_d r_{\pm})}{r_{\pm}^2-Q^2}$ & $\frac{- 2\pi (1-2\sigma_q r_{\pm})(r_{\pm}-2\mathcal{M})}{\sigma_q}$\\

       $C_{+} C_{-}$ & $4\pi^2 \left(Q^2-\sigma_r \right)^2 \frac{\left[ (Q^2-\sigma_r)-{\cal M}^2 \right]}
{\left[4(Q^2-\sigma_r)-3{\cal M}^2\right]}$ & $\frac{4\pi^2 Q^4 \left[ 4Q^4-Q^2(2{\cal M}+\sigma_d)^2+\sigma_d^2 Q^2 \right]}{4Q^4-Q^2 (2{\cal M}+\sigma_d)^2}$ & $\frac{16\pi^2 \mathcal{M}^2(8\mathcal{M}\sigma-1)}{\sigma^2}$ \\

\hline
   \end{tabular}
  \caption{A comparison of thermodynamical parameters for RN-Radiation, RN-Dust and Schwarzschild-Quintessence black holes.}\label{1table}
\end{table}
Heat capacity would be positive if:\\
\textbf{Case $1$:}   $1-2\sigma_q r_{\pm}<0$ and $r_{\pm}-2\mathcal{M}>0$.\\
\textbf{Case $2$:}  $1-2\sigma_q r_{\pm}>0$ and $r_{\pm}-2\mathcal{M}<0$.\\
Heat Capacity is negative if\\
\textbf{Case $a$:}  $1-2\sigma_q r_{\pm}>0$ and $r_{\pm}-2\mathcal{M}>0$,\\
\textbf{Case $b$:}  $1-2\sigma_q r_{\pm}<0$ and $r_{\pm}-2\mathcal{M}<0$.\\
Behavior of the heat capacity given in Eq. (\ref{sM23}) is shown in Fig. (\ref{3M}) for $\mathcal{M}=1$ and $\sigma_d= 0.01$, heat capacity is negative for $2<r<50$, positive for $0<r<2$ and $r>50$,  it is zero at $r=2$ and $r=50$.
A comparison of all the parameters calculated for Kiselev solutions is shown in Table. (\ref{1table}).
\section{Charged Dilaton Black Hole}
The action for charged black hole in string theory is \cite{ga4091}:
\begin{equation}
\mathcal{S}= \int d^4x \sqrt{-g}[-R+2(\bigtriangledown \varphi)^2+ e^{-2 a \varphi}F^2],
\end{equation}where $F$ is the Maxwell field, $\varphi$ is the scalar field and
$a$ is an arbitrary parameter specifying the strength
of dilaton and the Maxwell
field's coupling.
We are going to derive the area product formula and entropy product formula for a spherically symmetric dilaton black hole \cite{ga4091} whose metric can be
written in Schwarzschild-like coordinates as:
\begin{eqnarray}
ds^2=-{\cal N}(r)dt^{2}+\frac{dr^{2}}{{\cal N}(r)}+ {{\cal R}(r)}^{2}\left(d\theta^{2}+\sin^{2}\theta d\phi^{2}\right)
,\label{sph}
\end{eqnarray}
where the function ${\cal N}(r)$ is defined by
\begin{eqnarray}
{\cal N}(r) &=& \left(1-\frac{r_{+}}{r}\right)\left(1-\frac{r_{-}}{r}\right)^{\frac{1-a^{2}}{1+a^{2}}},
\end{eqnarray}
and
\begin{eqnarray}
{\cal R}^{2}(r) &=& r^{2}\left(1-\frac{r_{-}}{r}\right)^{\frac{2a^{2}}{1+a^{2}}}.
\end{eqnarray}
In these equations, $r_{+}$ and  $r_{-}$  are constants, which are related to mass
and charge of the black hole as:
\begin{eqnarray}
{\cal M} &=& \frac{r_{+}}{2} + \left(\frac{1-a^{2}}{1+a^{2}}\right) \frac{r_{-}}{2} , ~~~ Q = \sqrt{\frac{r_{+}r_{-}}{1+a^{2}}},
\end{eqnarray}
where as usual ${\cal M}$ is mass of the black hole and $Q$ is electric charge of the black hole. It may
be noted that $Q$ and $a$ are positive.
The horizons of the black hole are determined by the function ${\cal N}(r)=0$ which yields
\begin{eqnarray}
r_{+} &=& {\cal M}+ \sqrt{{\cal M}^{2}-\left(\frac{2n}{1+n}\right)Q^{2}},\label{hocde} \\
r_{-} &=& \frac{1}{n}\left[{\cal M}+ \sqrt{{\cal M}^{2}-\left(\frac{2n}{1+n}\right)Q^{2}}\right],\label{hocdc}
\end{eqnarray}
and $n$ is defined by
\begin{eqnarray}
n &=& \frac{1-a^{2}}{1+a^{2}}.
\end{eqnarray}

Here $r_{+}$ and  $r_{-}$ are called event horizon (${\cal H}^+$)  or outer horizon and Cauchy horizon (${\cal H}^-$)  or
inner horizon respectively, and  $r_{+}=r_{-}$ or ${\cal M}^{2}=\left(\frac{1+n}{2}\right)Q^{2}$
corresponding to the extreme charged dilaton black hole.\\
{\bf Case I:}  When $a=0$ or $n=1$, the metric corresponds to RN black hole.\\
{\bf Case II:} When $a=1$ or $n=0$, the metric corresponds to Gibbons-Maeda-Garfinkle-Horowitz-Strominger (GMGHS) black hole.
The expressions for surface gravity of dilaton black hole at both the horizons (${\cal H}^{\pm}$) are
\begin{eqnarray}
{\kappa}_{+} &=& \frac{1}{2r_{+}}\left(\frac{r_{+}-r_{-}}{r_{+}}\right)^{n}
 \,\, \mbox{and} \, \, {\kappa}_{-} =  0  ~.\label{sgcd}
\end{eqnarray}
The  black hole temperature or Hawking temperature at ${\cal H}^\pm$ are
\begin{eqnarray}
T_{+} &=& \frac{1}{4\pi r_{+}}\left(\frac{r_{+}-r_{-}}{r_{+}}\right)^{n} \mbox{and} \nonumber\\
T_{-} &=&0
\end{eqnarray}
Areas of the horizons (${\cal H}^\pm$) are
\begin{eqnarray}
{\cal A}_{+} &=& 4\pi r_{+}^2 \left(\frac{r_{+}-r_{-}}{r_{+}}\right)^{1-n}, ~~~
\, \, {\cal A}_{-} = 0.\label{arcd}
\end{eqnarray}
Interestingly, the area of both the horizons go  to zero at the extremal limit ($r_{+}=r_{-}$) which is
quite different from the well known RN and Schwarzschild black hole. The other characteristics of this spacetime is
that there is a curvature singularity at $r=r_{-}$.
Now the entropies of both the horizons (${\cal H}^\pm$) are
\begin{eqnarray}
{\cal S}_{+} &=& \pi r_{+}^2 \left(\frac{r_{+}-r_{-}}{r_{+}}\right)^{1-n}, ~~~ {\cal S}_{-}= 0.\label{etpcd}
\end{eqnarray}
Finally, the Komar energy is given by
\begin{eqnarray}
 E_{+} &=& \frac{r_{+}-r_{-}}{2} ~~~,   E_{-}=  0.\label{etpcd1}
\end{eqnarray}
Now we compute products of all the parameters given above:
\begin{eqnarray}
\mathcal{A}_{+}\mathcal{A}_{-} = 0, \\
\mathcal{S}_+\mathcal{S}_-=0,\\
\kappa_+\kappa_-= 0,\\
T_+ T_- = 0,\\
E_+E_-=  0.
\end{eqnarray}


Interestingly their products go to zero value and independent of mass thus are
universal quantities.
All of the above thermodynamical quantities must satisfied the first law of thermodynamics:
\begin{equation}\label{d10}
d{\cal M}=\mathcal{T}_{\pm} d\mathcal{A}_{\pm} +\Phi_{\pm}dQ,
\end{equation}
where
\begin{eqnarray}\label{d11}
\Phi_{\pm} &=& \Big(\frac{2n}{1+n}\Big)\frac{Q}{r_{\pm}}.
\end{eqnarray}
\begin{figure}
\centering
\includegraphics [width=9cm]{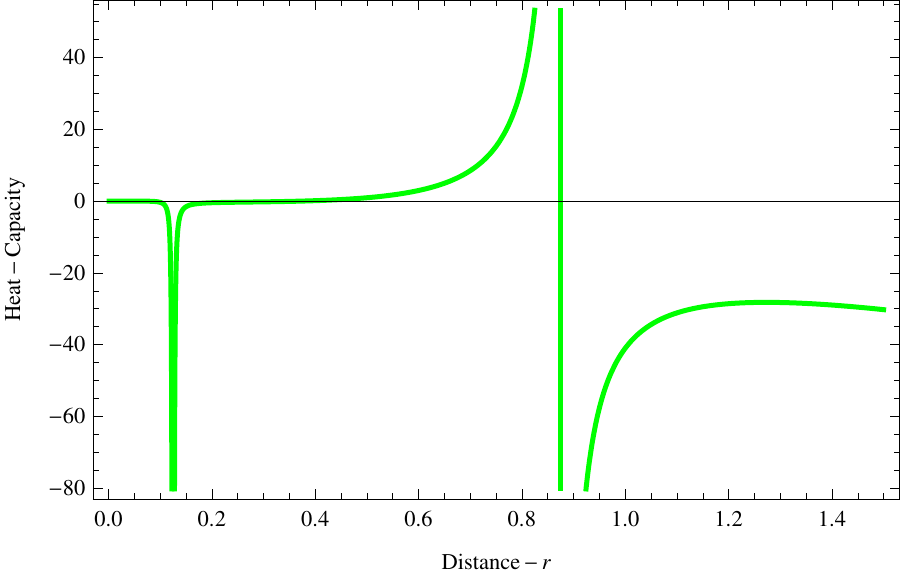}\caption{Heat capacity undergoes phase transition from instability to stability region, and again to instability region, with divergence at $r= 2Q^2/(1+n)$ and $2(1+n)Q^2/(1+n)$,  we chose $Q=0.5$, $n=3$ and ${\cal M}=1$.}\label{d3M}\end{figure}
The irreducible mass at $\mathcal{H}^{\pm}$ for this black hole is
\begin{eqnarray}
{\cal M}_{irr+} &=& \frac{r_{+}}{2} \left(\frac{r_{+}-r_{-}}{r_{+}}\right)^{\frac{1-n}{2}} ,~~~  {\cal M}_{irr-}=  0.\label{irrd}
\end{eqnarray}
Their product yields
\begin{eqnarray}\label{d2}
{\cal M}_{irr+} {\cal M}_{irr-} &=& 0.
\end{eqnarray}
The heat capacity for this dilaton black hole is calculated to be
\begin{eqnarray}
C_{+} &=& -2\pi r_{+}^2 \frac{\left[1-\frac{2n}{1+n}\frac{Q^2}{r_{+}}\right]\left[1-\frac{2}{1+n}\frac{Q^2}{r_{+}}\right]}
{\left[1-\frac{2(1+2n)}{1+n}\frac{Q^2}{r_{+}}\right]\left[1-\frac{2}{1+n}\frac{Q^2}{r_{+}}\right]^{n}}.~\label{cv1}
\end{eqnarray}
Due to curvature singularity at $r=r_{-}$ the heat capacity at the Cauchy horizon diverges.
Thus the product of heat capacity diverges.
Note that for odd $n$, heat capacity would be positive if
 \begin{eqnarray}r<\frac{2(1+2n)Q^2}{1+n}~~~~\text{and}~~ r>\frac{2 n Q^2}{1+n},
\end{eqnarray} or \begin{eqnarray}r>\frac{2(1+2n)Q^2}{1+n}~~~~\text{and}~~ r<\frac{2 n Q^2}{1+n},
\end{eqnarray}
and heat capacity is negative for:
\begin{eqnarray}r>\frac{2(1+2n)Q^2}{1+n}~~\text{and}~~ r>\frac{2 n Q^2}{1+n},\end{eqnarray}
or \begin{eqnarray}
r<\frac{2(1+2n)Q^2}{1+n}~~\text{and}~~ r<\frac{2nQ^2}{1+n},
\end{eqnarray}
behavior of the heat capacity  is shown in Fig. (\ref{d3M}). For $Q=0.5$ and $n=3$, given expression of heat capacity (Eq. (\ref{cv1})) diverges at $r=0.125$ and $r=0.875$, positive for $0.375<r<0.875$ and negative for $<0<r<0.125$, $0.125<r<0.375$, and $0.875<r< 1.5$ (for $r\in (0,1.5)$).
%
%

\section{Conclusion}
We have studied the thermodynamical properties on the inner and outer horizons of Kiselev solutions (RN black hole surrounded
by energy-matter (radiation and dust) and Schwarzschild black hole surrounded by quintessence) and charged dilaton black hole.
 We have studied some important parameters of black hole thermodynamics
with reference to their event and Cauchy horizons. We derive the expressions for temperatures and heat capacities
of all the black holes mentioned above.  It is observed that the product of
surface gravities, surface temperature product and product of
Komar energies at the horizons are not universal quantities for the Kiselev's solutions while products of areas and
entropies at both the horizons are independent of mass of black hole (except for Schwarzschild black hole surrounded by quintessence). For dilaton black hole these
products are universal except the products of specific heat which has shown divergent properties due to
the curvature singularity at the Cauchy horizon. Thus the implication of
these thermodynamical products may somehow give us further understanding of the microscopic nature of black hole
entropy (both exterior and interior) in the black hole physics.

Using the heat capacity expressions, stability regions of the black holes
 are also observed graphically. Figs. (\ref{1M}-\ref{d3M}) show that the above mentioned black holes undergo a phase transition under certain conditions on $r$. For RN black hole surrounded by radiation and dust and Schwarzschild black hole surrounded by quintessence, we derived the first law of thermodynamics using the Smarr formula approach. It is observed that third law of thermodynamics which states that "surface gravity, $\kappa$, of a black hole can not be reduced to zero in a finite sequence of processes" holds for all the above mentioned black holes. The derived expressions of $\kappa$ show that it is zero for extreme black holes only.


\begin{thebibliography}{99}
\bibitem{ch9670} D. Christodoulou, \textit{Phys. Rev. Lett.} {\bf 25}, 1596 (1970).
\bibitem{ch5271}  D. Christodoulou and R. Ruffini, \textit{Phys. Rev.} { \bf D 4}, 3552 (1971).
\bibitem{pe7771} R. Penrose and R. M. Floyd, \textit{ Nature} {\bf 229}, 177 (1971);
S. Hawking, \textit{Phys. Rev. Lett.} {\bf 26}, 1344 (1971).
\bibitem{be3373} J. D. Bekenstein, \textit{Phys. Rev.} {\bf D 7}, 2333 (1973);
J. D. Bekenstein, \textit{Phys. Rev.} {\bf D 9}, 3292 (1974).
\bibitem{ja6111}  M. Jamil, M. Akbar, \textit{Gen.Rel.Grav.} {\bf43}, 1061 (2011); M. Jamil, I. Hussain, M. U. Farooq, \textit{Astrophys. Space Sci.} {\bf335}, 339 (2011);  M. Jamil, I. Hussain, \textit{Int. J. Theor. Phys.} {\bf50}, 465 (2011);  M. Akbar, A. A. Siddiqui, \textit{Phys. Lett.} {\bf B  656}, 217 (2007).
\bibitem{ca0812}A. Castro and M. J. Rodriguez, \textit{ Phys. Rev.} {\bf D 86}, {024008} (2012).
    \bibitem{wa3114} J.~Wang, W.~Xu and X.~Meng, \textit{J. High Energy Phys.} {\bf 01}, 031 (2014).
\bibitem{cu6279} A. Curir, \textit{ Nuovo Cimento.} {\bf B 51}, {262} (1979).
\bibitem{pa9590} D. Pavon, \textit{Phys. Rev.} {\bf D 43},  2495 (1990);
S. W. Hawking, D. N. Page, \textit{Commun. Math. Phys.} {\bf 87},  577 (1983).
\bibitem {ne1865} E. T. Newman, E. Couch, K. Chinnapared, A. Exton,
A. Prakash, and R. Torrence, \textit{J. Math. Phys.} {\bf 6}, 918
(1965).
\bibitem{pr8714} P. Pradhan,  \textit{Eur. Phy. J.} { \bf C 74}, {2887} (2014).
\bibitem{prarx14} P. Pradhan, arXiv: 1503.04514 [gr-qc].
\bibitem{bushra} B. Majeed, M. Jamil,  arXiv:1507.01547 [hep-th].
    \bibitem{cv0111} M. Cvetic, G. W. Gibbons and C. N. Pope, \textit{Phys. Rev. Lett.} {\bf 106}, {121301} (2011).
\bibitem{vi1413}  M. Visser,  \textit{Phys. Rev.} {\bf D 88}, {044014} (2013).
\bibitem{ki8703} V. V. Kiselev, \textit{Class. Quant. Grav.} {\bf 20}, 1187 (2003).
    \bibitem{ja2415} M. Jamil, S. Hussain, and B. Majeed, \textit{Eur. Phys. J.} { \bf C 75}, 24 (2015).
\bibitem{ch83} S. Chandrashekar, \textit{The Mathematical Theory of Black Holes}, Clarendon Press, Oxford
(1983).
\bibitem{ha4471}S. Hawking, \textit{Phys. Rev. Lett.} {\bf 26}, 1344 (1971).
\bibitem{po02} E. Poisson,  \textit{A Relativist's Toolkit: The Mathematics of Black Hole Mechanics}, Cambridge University Press, (2007).
\bibitem{ko3459} A. Komar, \textit{Phys. Rev.} {\bf 113}, 934 (1959).
\bibitem{sm7173} L. Smarr, \textit{Phys. Rev. Lett.} {\bf 30},  71 (1973); L. Smarr, \textit{Phys. Rev.} {\bf D 7}, 289 (1973).
\bibitem{ba6173} J. M. Bardeen, B. Carter, S. W. Hawking,  \textit{Commun. Math. Phys.} {\bf 31}, 161 (1973).
\bibitem{gr3082}  D. J. Gross, M. J. Perry, and L. G. Yaffe, \textit{ Phys. Rev.} {\bf D 25}, 330 (1982).
\bibitem{ga4091} D. Garfinkle, G. T. Horowitz, A. Strominger, \textit{ Phys. Rev.} {\bf D 43}, 3140 (1991).


%
%
%
%
%
%
%
%
%
%
%




%
%

%
%
%
%

%
%
%
%
%
%
%
%
%
%
%
%
%
%

\end{thebibliography}
\end{document}